# Interplay of cascaded Raman- and Brillouin-like scattering in nanostructured optical waveguides

R. E. Noskov[1,2], J. R. Koehler[2] and A. A. Sukhorukov[3,2]

[1]*'Dynamics of Nanostructures' Laboratory, Department of Electrical Engineering,*
*Tel Aviv University, Ramat Aviv, Tel Aviv 6139001, Israel*
[2]*Max-Planck Institute for the Science of Light,*
*Staudtstr. 2, 91058 Erlangen, Germany*
[3]*Nonlinear Physics Centre, Research School of Physics and Engineering,*
*Australian National University, Canberra, ACT 2601, Australia*

**Abstract:** We formulate a generic concept of engineering optical modes and mechanical resonances in a pair of optically-coupled light-guiding membranes for achieving cascaded light scattering to multiple Stokes and anti-Stokes orders. By utilizing the light pressure exerted on the webs and their induced flexural vibrations, featuring flat phonon dispersion curve with a non-zero cut-off frequency, we show how to realize exact phase-matching between multiple successive optical side-bands. We predict continuous-wave generation of frequency combs for fundamental and high-order optical modes mediated via backward- and forward-propagating phonons, accompanied by periodic reversal of the energy flow between mechanical and optical modes without using any kind of cavity. These results reveal new possibilities for tailoring light-sound interactions through simultaneous Raman-like intramodal and Brillouin-like intermodal scattering processes.

Subwavelength confinement of light and sound in optical waveguides and cavities, achieved by precise structural engineering, allows optomechanical nonlinearities to surpass all other nonlinear effects[1,2,3,4,5]. This has led to the development of unprecedentedly efficient hybrid photonic-phononic signal processing devices, such as photonic radio-frequency notch[6–9] and multi-pole bandpass filters[10], novel sources for optical frequency combs[11–13] and lasers with ultra-low phase-noise[14]. These devices exploit an effect commonly known as stimulated Brillouin scattering (SBS): a refractive index grating provided by a propagating acoustic wave reflects pump photons into a counter-propagating Stokes wave at a frequency red-shifted by a few GHz. Its beat-note with the remaining pump signal reinforces, in turn, the acoustic wave via electrostriction or radiation pressure[15].

Strong light-sound interaction in nanostructured waveguides also allowed SBS-like processes to be enriched with two more scattering phenomena involving two co-propagating optical modes: stimulated inter-polarization scattering (SIPS) between two orthogonal polarization directions[16], and stimulated inter-modal scattering (SIMS) between co-polarized optical modes[17,18]. Much like in SBS, the frequency shifts in both SIPS and SIMS depend on the pump laser frequency and both types of transitions require phonons with a significant amount of momentum: as a result, each phonon mediates only one specific transition. Although favorable for all-optical reconfigurable isolation schemes[19], the large axial phonon wavevector makes cascaded stimulated scattering extremely inefficient[20]. So far, the reported demonstrations exploited SBS in optical cavities, where phase-matching of cascaded transitions with the same phonon is possible due to interplay with four-wave mixing[13].

A fundamentally different situation emerges for photons interacting with THz-frequency optical phonons associated with molecular vibrations (also referred to as coherence waves): their frequency-wavevector relation is, in strong contrast to the dispersion of acoustic modes mediating SBS, perfectly flat, so that both the optical pump wavelength and the phonon wavevector can be freely chosen without changing the phonon frequency. As its group velocity equals to zero, the single optical phonon can mediate transitions between pairs of adjacent side-bands within a given optical mode, their frequencies spaced by the common Raman shift. The number of side-bands generated in the optical spectrum via cascaded stimulated Raman scattering (SRS) is, however, restricted by two effects: on the one hand, Raman transitions experience strong dephasing due to group-velocity dispersion (GVD) increased with rising an order of the involved adjacent side-bands, rendering their excitation more and more inefficient. On the other hand, if pump-to-Stokes and pump-to-anti-Stokes transitions are driven by phonons of identical frequency and momentum (e.g. for negligible GVD) and if only single-frequency laser light is launched, each generated phonon would be immediately annihilated. This, in turn, prevents a phonon population from building up, leading to coherent gain suppression. First predicted[21] and observed[22] in free space, its dramatic impact on optical frequency comb generation has been recently studied in gas-filled hollow-core photonic crystal fibers[23,24].

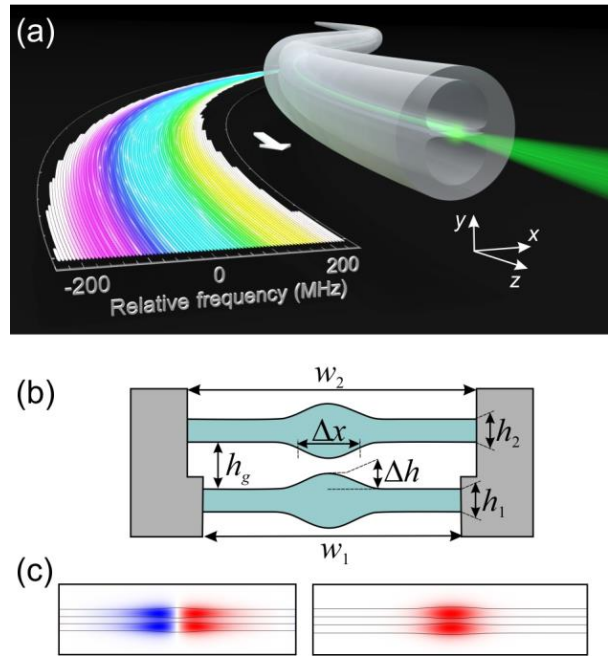

**Fig. 1: (a)** Artistic view of frequency comb generation in a dual-nanoweb fibre for stimulated Raman-like scattering. The beat-note of two-frequency laser light, launched at the fibre input, stimulates flexural vibrations which mediate cascaded transitions between many Stokes and anti-Stokes side-bands. **(b)** Sketch of a cross-section for an idealized dual-nanoweb structure: At their suspensions, the nanoweb thickness is $h_1 = h_2 = h$ and their gap $h_g = h = 500$ nm. A Gaussian thickness profile of the nanowebs is defined by a height $\Delta h = 100$ nm and a waist $\Delta x = 1.5$ μm. **(c)** Simulated electric field distribution for fast (left) and slow (right) optical modes.

Hence, to generate a large number of side-bands even in presence of significant GVD, the phonon frequency has to be orders of magnitude smaller than the optical pump frequency. Although impossible for Raman molecules, this condition is typically met for transverse acoustic resonances in nanostructured waveguides representing artificial Raman oscillators: the dispersion of their associated acoustic phonons is, similar to optical phonons in gases, flat close to cut-off allowing them to mediate intra-modal stimulated Raman-like scattering (SRLS) among many orders of Stokes and anti-Stokes side-bands[25]. Such optical frequency combs were generated for SRLS by GHz vibrations in the ~1 μm-thin solid core of a photonic crystal fibre (PCF)[20,25], a membrane-suspended silicon waveguide[26] and also by flexural vibrations at ~6 MHz in a dual-nanoweb fiber[27], the system investigated in this work. Its structure is formed by a capillary fibre that holds two flexible silica membranes ("nanowebs") in its hollow channel, spaced by a tiny interweb gap (Fig. 1). Transverse radiation pressure acting between these coupled waveguides gives considerable rise to the refractive indices of guided optical supermodes at mW launched optical powers[1,2]. Despite such giant optomechanical nonlinearity, this system features perfect conditions for gain suppression: the Raman-like frequency shift provided by flexural nanoweb vibrations (~6 MHz) is so small that GVD is fully negligible. Observing the nanowebs self-oscillate at their mechanical resonant frequency, when only a few mW of single-frequency laser light was launched, was therefore unexpected. This is because so far hardly any attention has been paid to the interplay of processes as diverse as Raman and Brillouin scattering, or their respective analogues SRLS and SIPS/SIMS in waveguides: they typically occur in fundamentally different frequency ranges, rendering them intrinsically uncoupled.

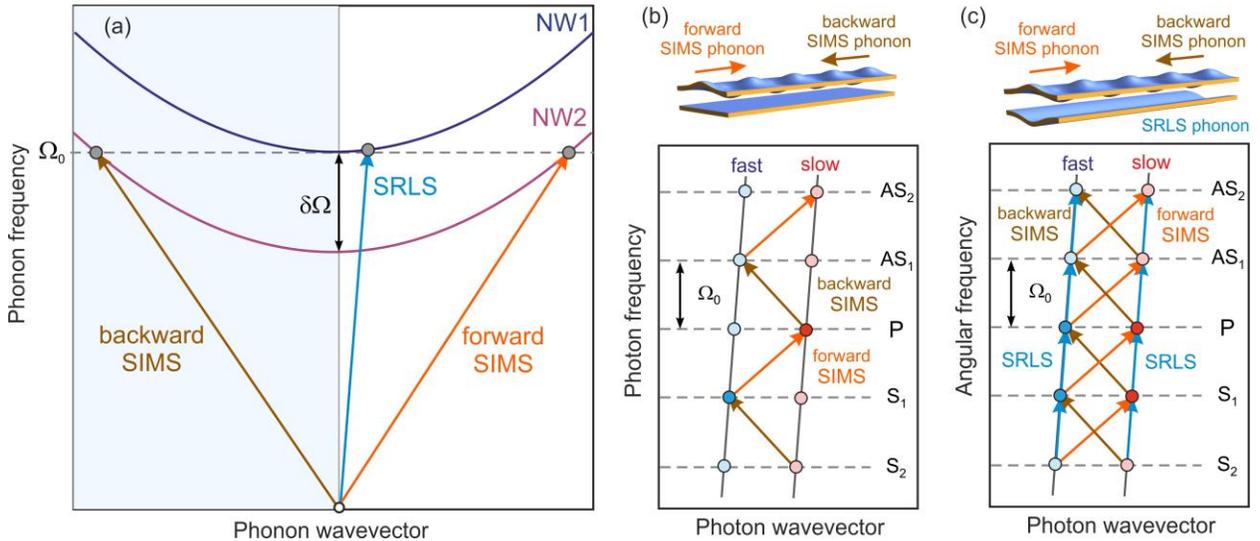

**Fig. 2:** **(a)** Dispersion diagram (frequency vs. wavevector) for phonons of the fundamental flexural resonance in each nanoweb (NW1 and NW2), the white circle indicating zero frequency/wavevector and arrows denoting SRLS (in blue) and forward/backward SIMS phonons (in orange/brown). Slightly different widths cause a frequency offset $\delta\Omega$ between their respective dispersion curves, allowing at a certain wavevector SRLS and SIMS to occur at the same frequency $\Omega_0$. **(b)-(c)** Optical

dispersion schemes including the fundamental (slow) and double-lobed (fast) optical modes (grey lines). Dark filled circles indicate optical side-bands seeded at the fibre input, images on top sketch characteristic nanoweb deflection profiles. (b) Cascade of forward and backward SIMS transitions, with even- and odd-order side-bands alternatingly created in the slow and fast optical modes. (c) Interplay of SRLS (blue) with forward/backward SIMS (orange and brown) when seeding the pump (P) and 1$^{st}$-order Stokes (S$_1$) side-band in both optical modes.

Based on a general theoretical model, this paper reveals conditions under which this paradigm breaks down: for this we consider an axially perfectly uniform dual-nanoweb fiber (cross-section sketched in Fig.1(b)) supporting one fundamental flexural mode in each nanoweb and also two co-polarized optical eigenmodes referred to as slow (*s*) and fast (*f*). On the one hand, a highly unusual situation stems from the simultaneous presence of a forward- and a backward-propagating flexural wave at the same frequency: these can alternatingly mediate $s \rightarrow f$ and $f \rightarrow s$ transitions in a ladder-like fashion (Figs. 2(a) and (b)), leading, in strong contrast to SBS, to a unique cascade of transitions between numerous exactly phase-matched SIMS side-bands. The optical frequency comb generated this way will feature all its even-order side-bands (spaced by twice the SIMS frequency shift) in the fundamental optical mode, whereas odd-order ones will appear in the higher-order optical mode. On the other hand, intricate co-operation between Raman-like SRLS and Brillouin-like SIMS becomes possible (Fig. 2(c)), a fact that has been found to frustrate SRLS gain suppression in a particular dual-nanoweb sample with strong axial non-uniformities[18]. We show that SRLS and SIMS can be synchronized (i.e. happen at the same frequency) even in the absence of any axial non-uniformities: the only requirements are a slight offset between the flexural frequencies in each nanoweb as well as some tuning of the optical pump wavelength. The resulting evolution of the optical side-band powers along the fibre shows that under these circumstances energy transfer between individual optical modes and flexural vibrations can go to-and-fro, making it possible to enhance one flexural mode via supression of another.

## RESULTS AND DISCUSSION

### Theory of stimulated light scattering by flexural vibrations

We demonstrate the two aforementioned effects in an idealized dual-nanoweb structure whose cross-section is sketched in Fig. 1(b): each silica nanoweb features a width $w_j$ ($j$ = 1,2 refers to upper and lower nanoweb). At their suspensions to the inner capillary walls (which are indicated in grey), the thicknesses of the nanowebs are $h_j$ and their spacing is $h_g$. For simplicity, we assume the following $h_1 = h_2 = h_g = h$ = 500 nm. The formation of bound optical modes guiding light in the coupled waveguides is ensured by a slightly convex design of both nanowebs: their thickness at the centre of the structure is 100 nm larger than at the edges and supposed to follow a Gaussian profile. Choosing its waist ~1.5 µm along with $w_j \geq$ 15 µm makes it possible to provide low losses for both the single-lobed slow (s) and double-lobed fast (f) optical modes (Fig. 1(c)). Assuming $h \ll w_j$, SIMS transitions between these two optical modes and as well as SRLS transitions within each of them are then mediated by phonons of the same (fundamental) flexural resonance.

These interactions can be described by a set of coupled-mode equations governing the slowly-varying electric field amplitudes and the envelope of the flexural vibrations. In the steady state ($\partial/\partial t$ = 0) they take the form (see Methods for derivation)

$$\frac{\alpha_s}{2} s_n + \frac{\partial s_n}{\partial z} = i \frac{\omega_n}{\omega_0} \sum_{j=1}^{2} \left[ \kappa_{sRs}^{(j)} \left( R_j s_{n-1} + R_j^* s_{n+1} \right) + \kappa_{sMf}^{(j)} \left( M_j^+ f_{n-1} + M_j^{-*} f_{n+1} \right) \right], \tag{1a}$$

$$\frac{\alpha_f}{2} f_n + \frac{\partial f_n}{\partial z} = i \frac{\omega_n}{\omega_0} \sum_{j=1}^{2} \left[ \kappa_{fRf}^{(j)} \left( R_j f_{n-1} + R_j^* f_{n+1} \right) + \kappa_{fMs}^{(j)} \left( M_j^+ s_{n-1} + M_j^{-*} s_{n+1} \right) \right], \tag{1b}$$

$$R_j \frac{i\Omega \Gamma_{jR} + \Omega^2 - \Omega_{jR}^2}{2i\Omega} = i\kappa_{Rss}^{(j)} \Phi_{ss} + i\kappa_{Rff}^{(j)} \Phi_{ff}, \tag{1c}$$

$$M_j^+ \frac{i\Omega \Gamma_{jM} + \Omega^2 - \Omega_{jM}^2}{2i\Omega} + V_{jM} \frac{\partial M_j^+}{\partial z} = i\kappa_{Msf}^{(j)} \Phi_{sf}, \tag{1d}$$

$$M_j^- \frac{i\Omega \Gamma_{jM} + \Omega^2 - \Omega_{jM}^2}{2i\Omega} - V_{jM} \frac{\partial M_j^-}{\partial z} = i\kappa_{Msf}^{(j)} \Phi_{fs}. \tag{1e}$$

Here, $s_n$ and $f_n$ represent the slowly-varying electric field amplitudes of the $n^{\text{th}}$ side-band in the slow and fast mode, respectively, $\alpha_s$ and $\alpha_f$ are the loss coefficients of the slow and fast optical mode, $R_j$ the slowly-varying amplitudes of SRLS phonons in nanoweb $j$, $M_j^+$ and $M_j^-$ the amplitudes of forward- and backward-propagating SIMS phonons, $\Gamma_{jR}$ and $\Gamma_{jM}$ their Brillouin linewidths, and $V_{jM}$ the group velocity of SIMS phonons. The parameters $\kappa_{mRm}^{(j)}, \kappa_{mMm}^{(j)}, \kappa_{Rss}^{(j)}, \kappa_{Rff}^{(j)}, \kappa_{Msf}^{(j)}$ are the optoacoustic coupling rates (see Methods for detailed definitions of all these quantities).

It is instructive to note that the Manley-Rowe relations require that, when the optical losses are negligible ($\alpha_j = 0$), the rate of pump photon destruction equals to the rate of Stokes and anti-Stokes photons and phonons creation. This pertains coupling rates as $\omega_0 \kappa_{Rss}^{(j)} = \Omega c \kappa_{sRs}^{(j)} = \Omega c \kappa_{fRf}^{(j)}$ and $\omega_0 \kappa_{Msf}^{(j)} = \Omega V_{jM} \kappa_{sMf}^{(j)} = \Omega V_{jM} \kappa_{fMs}^{(j)}$, manifesting the existence of a single coupling coefficient that captures all reversible optical forces and scattering for a given process[28,29]. The coupling rates are related to the Brillouin gain factors via $g_{mm}^{(j)} = 4\kappa_{mRm}^{(j)} \kappa_{Rmm}^{(j)} / \Gamma_{jR}$ ($m = s$ or $f$), $g_{sf}^{(j)} = 4\kappa_{sMf}^{(j)} \kappa_{Msf}^{(j)} / \Gamma_{jM}$ and $g_{fs}^{(j)} = 4\kappa_{fMs}^{(j)} \kappa_{Msf}^{(j)} / \Gamma_{jM}$ [30]. For the structure under investigation, simulations yield $g_{ss}^{(j)} \approx 0.174 \ \mu\text{m}^{-1}\text{W}^{-1}$, $g_{ff}^{(j)} \approx 0.133 \ \mu\text{m}^{-1}\text{W}^{-1}$, and $g_{sf}^{(j)} \approx g_{fs}^{(j)} \approx 0.051 \ \mu\text{m}^{-1}\text{W}^{-1}$. These are unprecedentedly large values compared to conventional SBS in fibers $\sim 10^{-9} \ \mu\text{m}^{-1}\text{W}^{-1}$ [31] and silicon waveguides $\sim 10^{-4} \ \mu\text{m}^{-1}\text{W}^{-1}$ [32] as well as SRLS in solid-core PCFs $\sim 10^{-6} \ \mu\text{m}^{-1}\text{W}^{-1}$ [20], achieved in the absence of any optical or acoustic cavity, resulting from long phonon lifetimes and tight optomechanical confinement, allowing us to consider optoacoustic scattering in isolation from all other nonlinear processes.

The quantities $\Phi_{ss} = \sum_n s_n s_{n-1}^*$ and $\Phi_{ff} = \sum_n f_n f_{n-1}^*$ represent the summed-up optical beat-notes of all pairs of adjacent side-bands in the slow and fast optical mode, respectively, and can be viewed as a generalized optical pressure in the equation-of-motion (1c) for SRLS vibrations.

Similarly, $\Phi_{sf} = \sum_n s_n f_{n-1}^*$ and $\Phi_{fs} = \sum_n s_n^* f_{n+1}$ are the corresponding beat-note pressures driving forward- and backward-propagating SIMS phonons, respectively.

When two-frequency laser light is launched into the fibre so as to seed the pump and first-order Stokes side-bands simultaneously in both optical modes, the general boundary conditions of Eqs.(1) are

$$s_n(0) = \sqrt{P_{s,0}/P_0}\,\delta_{n,0} + \sqrt{P_{s,-1}/P_0}\,\delta_{n,-1} + \sqrt{P_{noise}/P_0}\,(\delta_{n,-2} + \delta_{n,+1}),$$
$$f_n(0) = \sqrt{P_{f,0}/P_0}\,\delta_{n,0} + \sqrt{P_{f,-1}/P_0}\,\delta_{n,-1} + \sqrt{P_{noise}/P_0}\,(\delta_{n,-2} + \delta_{n,+1}),\ M_j^+(0) = M_j^-(L) = 0, \quad (2)$$

where $\delta_{n,m}$ is the Kronecker delta ($m = -2, -1, 0, +1$), $L$ the fibre length, $P_{s,0(-1)}$ and $P_{f,0(-1)}$ the powers launched into the slow and the fast optical modes at the pump and the 1$^{st}$-order Stokes frequencies ($P_0 = P_{s,0} + P_{s,-1} + P_{f,0} + P_{f,-1}$). The effective noise power $P_{noise} = k_B T \omega_0 \Gamma_{jK}/(2\Omega_{jK})$ ($K = R$ or $M$) in optical side-bands adjacent to the seeded ones accounts for the large population of thermal phonons available for spontaneous scattering in the system[31] ($k_B$ is Boltzmann's constant and $T$ is the ambient temperature). When the backward SIMS phonon is included in Eqs. (1), the boundary conditions are split which means that one has to solve coupled dynamical equations describing processes evolving in opposite directions. To keep causality, an iterative scheme was used in which Eqs. (1a) – (1d) and Eq. (1e) were solved sequentially until convergence was reached.

**Intricate coupling of forward and backward inter-modal scattering**

We begin our analysis with a situation in which Brillouin-like scattering (SIMS) is fully uncoupled from Raman-like scattering (SRLS), while forward- and backward-propagating flexural waves are permitted to coexist at the same frequency. In an idealized dual-nanoweb fibre this generally happens if the pump wavelength is not explicitly tuned to fulfill conditions for the synchronization of SRLS and SIMS (see next Section). Provided that the pump side-band (P) in the slow and the 1$^{st}$-order Stokes side-band (S1) in the fast optical mode are seeded, a cascade of SIMS transitions is driven. The boundary conditions (Eq.(2)) can be recast for this situation with $P_{s,+1} = P_{s,-1} = P_{noise}$ and $P_{f,0} = P_{f,-2} = P_{noise}$ as $s_n(0) = \sqrt{P_{s,0}/P_0}\,\delta_{n,0} + \sqrt{P_{noise}/P_0}\,(\delta_{n,-1} + \delta_{n,+1})$ and $f_n(0) = \sqrt{P_{f,-1}/P_0}\,\delta_{n,-1} + \sqrt{P_{noise}/P_0}\,(\delta_{n,0} + \delta_{n,-2})$. For the sake of simplicity, we also accept $V_j^M = 0$ and $\Omega = \Omega_{jM}$ (effects induced by non-zero $V_j^M$ will be discussed below). The optical pressures $\Phi_{sf}$ and $\Phi_{fs}$ driving $s \to f$ and $f \to s$ SIMS transitions evolve then according to the relations

$$\frac{\partial |\Phi_{sf}|^2}{\partial z} = \left[-\alpha_s - \alpha_f + \Delta P(z)\left(g_{sf}^{(1)} + g_{sf}^{(2)}\right)\right] |\Phi_{sf}|^2,$$

$$\frac{\partial |\Phi_{fs}|^2}{\partial z} = \left[-\alpha_s - \alpha_f - \Delta P(z)\left(g_{fs}^{(1)} + g_{fs}^{(2)}\right)\right] |\Phi_{fs}|^2,$$

where $\Delta P(z) = P_0 \sum_n \left( |s_n(z)|^2 - |f_n(z)|^2 \right)$ is the contrast of optical power between the slow and the fast modes. Thus, $\Delta P$ governs the stimulation of $s \to f$ and $f \to s$ SIMS transitions: if the input power in the slow mode is greater than in the fast one, $\Phi_{sf}$ grows until reaching its maximum for $\Delta P(z) = 0$, when the generation of forward SIMS phonons driving the $P \to S_1$ transition is strongest. After that, $\Delta P$ changes its sign, causing decay of $\Phi_{sf}$ and rise of $\Phi_{fs}$ which results in noise-driven creation of backward SIMS phonons stimulating the $S_1 \to S_2$ transition. Next, $s \to f$ and $f \to s$ transitions to higher anti-Stokes and Stokes orders follow one by another in a ladder-like fashion. Thus, an optical frequency comb is generated, with even- and odd-order side-bands appearing alternatingly in the slow and the fast mode (see dispersion diagram in Fig. 2(b)).

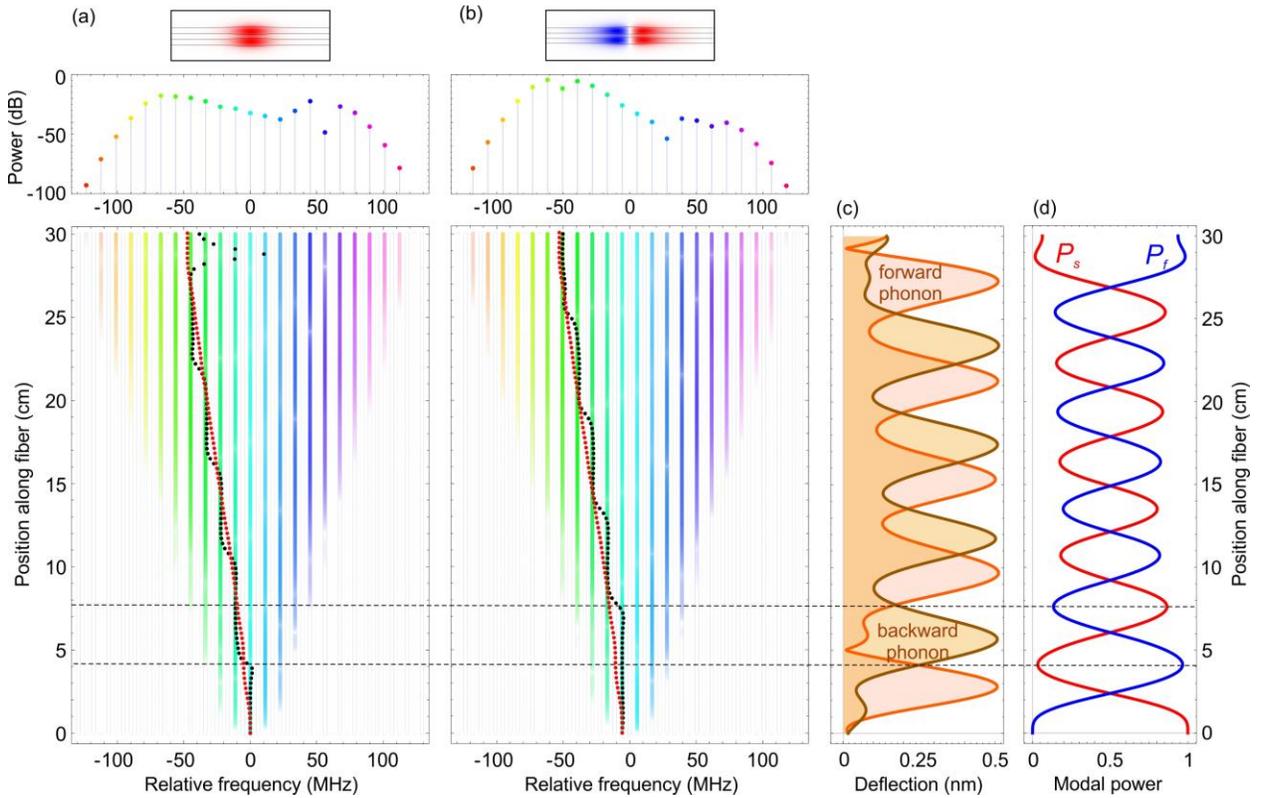

**Fig. 3:** Evolution of power in side-bands of the optical frequency combs along the fiber in the **(a)** slow and **(b)** fast optical modes, generated by a cascade of alternating forward/backward SIMS transitions (see Fig. 2(b) for the dispersion diagram). The color of the plotted lines represents the respective side-band order, brightness indicating their normalized powers in dB. Black dotted curves represent the mean comb frequencies separately for each mode; while the red dotted curve shows the mean frequency of the full optical spectrum. Shown on top are the optical spectra at the fibre output: even-order side-bands are generated in the slow and odd-order ones in the fast mode. **(c)** Nanoweb peak deflection caused by forward (orange) and backward (brown) SIMS phonons and **(d)** fraction of optical power guided in the slow (red) and fast (blue) optical modes, plotted versus position along the fibre. Dashed lines indicate that the maximal contrast in modal powers corresponds to the steps of comb mean frequencies.

Figures 3 (a) and (b) plot the evolution of the side-band powers along the fibre in the slow and the fast mode at an optical pump wavelength $\lambda = 1.5$ μm and for $w_1 = 23$ μm, $w_2 = 28.06$ μm, $P_0 = 3.5$ mW, $P_{s,0}/P_0 = 0.99$, $P_{f,-1}/P_0 = 0.01$, and $\alpha_s = \alpha_f = 0$. The associated slowly-varying envelopes of forward and backward SIMS phonons propagating with $V_{jM} = 12$ m/s and $\Omega_0 = 2\pi \times 5.6$ MHz are shown in Fig. 3 (c). The mean frequency of the generated frequency comb relative to the pump side-band, calculated by $\Omega_s(z) = \sum_n n\Omega_0 |s_n(z)|^2 / \sum_n |s_n(z)|^2$ in the slow and $\Omega_f(z) = \sum_n n\Omega_0 |f_n(z)|^2 / \sum_n |f_n(z)|^2$ in the fast mode, is plotted as a dotted black line, a net downshift (red dotted curve) indicating mechanical work by the optical driving field over acoustic vibrations. As can be seen in Figs. 3(a) and (b), the profile of the mean comb frequency contains a periodic sequence of steps: these indicate regions of maximal work that coincide with peaks in the modal power contrast (plotted in Fig. 3 (d)).

On the contrary, close-to-zero (or zero when $V_j^M \sim 0$) modal power contrast $\Delta P$ results in the highest vibrational amplitudes (compare Figs 3 (c) and (d)). The resulting pattern of acoustic vibrations contains quasi-periodic fringes of coupled forward- and backward-propagating phonons. The remarkably high SIMS gain permits tuning of the grating period ranging from several millimeters for $P_0 \sim 1$ W to meter-scale for $P_0 < 1$ mW which can find applications in microwave photonics[15].

It is instructive to note that the slight blue-shifts of the mean comb frequency in some fibre sections are caused by nonlocality of the optomechanical interactions, allowing a reversed energy flow from the mechanical vibrations back to the optical fields. Such a situation emerges as a result of growing phase delay between optical pressure $\Phi_{fs}$ and backward SIMS phonons as they travel back from the point where they were generated. The scale of this nonlocality is determined by the phonon propagation distance given by $V_{jM}\tau_M$, where $\tau_M = 1/\Gamma_{jM}$ is the lifetime for SIMS phonons. For the considered structure $V_{jM} = 12$ m/s and $\tau_M = 250$ μs yield the phonon propagation distance ~3 mm[18].

Another remarkable effect is straight coupling between forward- and backward-propagating SIMS phonons through reflection from the fiber end faces. This condition impacts the phase relationship between acoustic vibrations and optical modes close to the fiber output which resulted in suppression of the backward SIMS phonon and the associated upshift of the mean comb frequency for the slow mode at the ~28 cm position (see Figs. 3(a) and (c)).

**Cascaded inter- and intra-modal scattering**

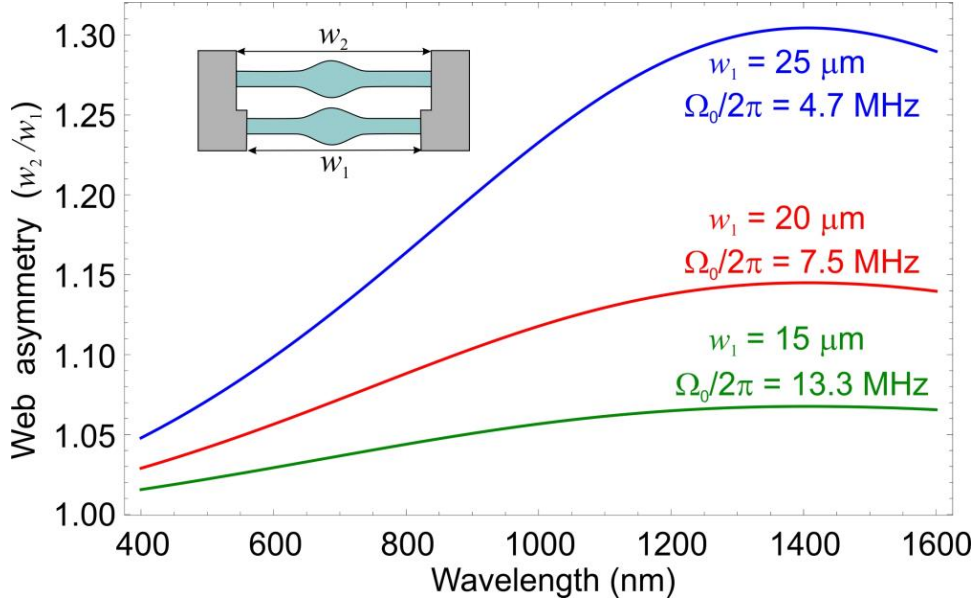

**Fig. 4:** Ratio of the web widths needed to synchronize SRLS and SIMS versus optical wavelength. In the simulations, the width of the upper web was varied, while assuming a constant width of the lower nanoweb of 15 μm (green line), 20 μm (red) and 25 μm (blue).

Typically, the frequency shift in SIMS transitions is larger than for SRLS because the flexural dispersion of the fundamental resonance in a single nanoweb is monotonous (see Fig. 2(a)). Hence, to study the interplay of SRLS and SIMS at a common frequency, we tailor the flexural dispersion curves of the nanowebs to achieve a frequency offset allowing both SRLS and SIMS with a single Stokes frequency shift. This requires geometrical asymmetry, i.e. different widths of upper and lower nanowebs, and, on the other hand, some tuning of the optical pump wavelength. We simulate the optical and acoustic dispersions (see Methods) and plot in Fig. 4 the structural asymmetry required to synchronize SRLS and SIMS, i.e. the ratio $w_1/w_2$, as a function of the pump wavelength for three structures featuring different SRLS phonon frequencies $\Omega_0$. Notably all these curves are non-monotonous, which means that for any given structure synchronized SRLS and SIMS may occur at up to two specific optical wavelengths.

SRLS features several key properties that were characterized in detail in Refs. [18] and [20]: to unbalance gain suppression, it generally requires seeding the pump and the Stokes side-bands within one optical mode. The optical pressures $\Phi_{ss}$ and $\Phi_{ff}$ driving flexural vibrations close to cut-off (see Fig. 2(a)) are in the absence of optical losses solely defined by the input optical field. Unaffected by power redistribution between the side-bands, the fibre length, and the amplitude of flexural vibrations[33], they are constant along the whole waveguide. This means that the work done by the optical field over them is also unvaried along the fibre.

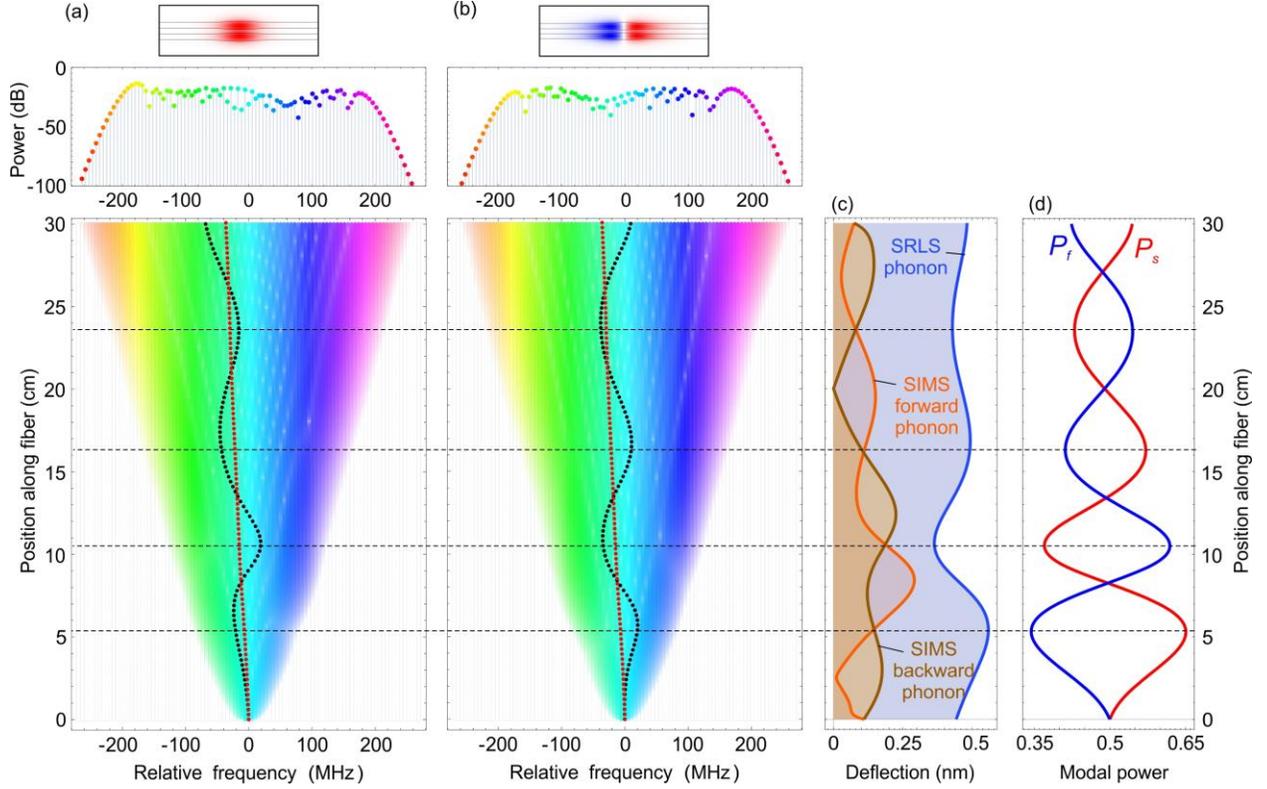

**Fig. 5:** Same plots as in Fig. 4 for an interplay between SRLS and forward/backward SIMS (see Fig. 2(c) for phase diagram). Under these circumstances, each optical side-band is linked to adjacent ones within the same mode by SRLS and in opposite modes by SIMS. Dashed lines denote correspondence between the highest contrasts in modal powers and turning points of comb mean frequencies.

Next, we consider the interplay of SRLS and forward/backward SIMS which occurs when the pump and 1$^{st}$-order Stokes side-bands are seeded in both optical modes at the same time. Under these circumstances, all possible transitions are simultaneously driven, leading to the generation of frequency combs in each optical mode with all side-bands populated (see Fig. 2(c)). Figure 5 shows the spatial evolution of all side-band powers and vibrational amplitudes along the fibre at an optical pump wavelength $\lambda = 1.55$ µm and for $w_1 = 23$ µm, $w_2 = 28.06$ µm, $P_0 = 4$ mW, $P_{s,0}/P_0 = P_{f,0}/P_0 = 0.49$, $P_{s,-1}/P_0 = P_{f,-1}/P_0 = 0.01$, $\alpha_s = 0.01$ m$^{-1}$ and $\alpha_f = 0.1$ m$^{-1}$. While isolated SRLS spreads the power among side-bands within a given optical mode by allowing the optical field to do constant mechanical work over flexural vibrations along the fiber[18], and pure SIMS makes optical modes to exchange power by performing mechanical work periodically one after another (Figs. 3 (a) and (b)), their coupling yields surprising results: the spatial evolution of the comb mean frequencies features oscillations with pronounced amplitudes covering ~10 sidebands, occurring in opposite phase for the slow and the fast modes (Figs. 5(a) and (b)). This indicates that energy transfer between individual optical modes and vibrations can go to-and-fro without using any form of cavity, in contrast to energy flow between electromagnetic and acoustic forms which is always one-way, as evidenced by a net downshift of the optical spectrum mean frequency. Seeded by the light launched at the input fibre face end, SRLS creates many side-bands in each optical mode.

However, the phase-lags between SRLS phonons and optical side-bands are not fixed anymore, as it was in case of isolated SRLS[33], since each pair of adjacent side-bands in both optical modes drives SIMS transitions at every position along the fibre (Fig. 2(c)). Hence, SRLS and forward/backward SIMS phonons can be partially absorbed by side-bands of slow or fast mode depending on respective phase delay, enabling oscillations of comb mean frequencies.

Notably, the positions where the modal power contrast takes maximal values, indicating maximal work done by SIMS transitions, coincide with the turning points of comb mean frequencies as well as with maximal and minimal values of SRLS phonon amplitudes (compare Figs. 5(a)-(d)). This is because SIMS processes provide power transfer between optical modes mediated by generation/suppression of respective vibrations and interfere with SRLS.

Finally, we point out that the strength of these oscillations in powers between optical modes and mechanical resonances stem from the asymmetric contributions of the fundamental and high-order optical mode to the optomechanical interactions. Since GVDs, modal indices, and nonlinear optomechanical coupling rates were found to be very close for both modes (see Methods), the significant imbalance in the modal loss rates $\alpha_s$ and $\alpha_f$ is the main reason for the oscillations of the comb mean frequencies over ~10 sidebands in each optical mode. We note, however, that these oscillations do not disappear even if $\alpha_s$ and $\alpha_f$ are equal: driven by slight differences in other parameters, the oscillation amplitude will drop under these circumstances to ~2 side-bands.

**CONCLUSION**

We have shown that a pair of highly mechanically compliant light-guiding membranes provides a promising playground for the generation of unique optomechanical patterns without using any kind of optical or acoustic cavity. We derive a general theoretical model describing inter- and intra-modal light scattering to multipole Stokes and anti-Stokes orders. Stimulated Raman-like light scattering generates a frequency comb for a single optical mode associated with static flexural vibrations of a constant amplitude along the whole waveguide. Stimulated inter-modal light scattering creates frequency combs for fundamental (slow) and high-order (fast) optical modes at even and odd side-bands, respectively, backward- and forward-propagating flexural phonons forming an effective optoacoustic grating. Its period can be tuned over a wide range from mm- to meter-scale. Adjustment of nanoweb widths allows matching both these Raman-like and Brillouin-like processes. As a result, the system drives intricate optomechanical patterns permitting periodic reversal of the energy flow between mechanical and optical modes.

Our results can be transferred to a variety of systems supporting Raman-like and inter-modal scattering, providing great capacity for practical use. Specifically, in silicon photonics combination of SIMS and SRLS would enable novel ways of active mode conversion, which may have applications in switching, power routing and on-chip mode-division multiplexing[17,26,34,35]. The considered dual-web waveguides allow easy independent geometric tuning of the wavevectors and frequencies (up to a few GHz by using thicker and less-wide webs[17,26]) for SIMS and SRLS phonons without influence on optical modes, unlike in most Brillouin systems[3]. Therefore, interplay of SIMS and SRLS can be used as a flexible platform to generate acoustically decoupled but opto-mechanically related static and propagating-wave phonons for on-chip acousto-optic devices. Beyond that, our findings can be used for microwave-over-optical signal processing[36,37] and optical frequency metrology[38].

**ACKNOWLEDGEMENT**

A.A.S. gratefully acknowledges financial support from the Alexander von Humboldt Foundation and the Australian Research Council (Discovery Project No. DP130100086). The authors acknowledge Prof. P. St.J. Russell for fruitful discussions.

**METHODS**

**Derivation of optomechanical coupled-mode equations**

In general, the interaction between TE-polarized optical waves and flexural phonons can be described by the following equations[30,39,40]

$$D_j\left(1+\tau\frac{\partial}{\partial t}\right)\left(\frac{\partial^4}{\partial x^4}+2\frac{\partial^4}{\partial x^2 \partial z^2}+\frac{\partial^4}{\partial z^4}\right)\delta_j + \sigma_j \frac{\partial^2 \delta_j}{\partial t^2} = p_j^{opt}, \qquad (3)$$

$$\frac{\partial^2 E_x}{\partial z^2} - \frac{n_m^2}{c^2}\frac{\partial^2 E_x}{\partial t^2} = \frac{1}{\varepsilon_0 c^2}\frac{\partial^2 P_x^{NL}}{\partial t^2}, \qquad (4)$$

where $\delta_j$ is the web deflection, $p_j^{opt}$ the optical pressure, $n_m$ the optical modal index, $\varepsilon_0$ the dielectric permittivity and $c$ the speed of light in vacuum, $D = \Upsilon h^3/[12(1-\nu^2)]$ the nanoweb flexural rigidity, $\sigma = \rho h$ the mass of a nanoweb per unit area, $\rho$ the density of silica, $\Upsilon$ Young's modulus, $\nu$ Poisson's ratio, and the parameter $\tau$ describes material viscosity. From this point on we neglect spatial variations of the web widths in the flexural wave equation (3).

Dominating over electrostriction because of the small interweb gap, the optical gradient pressure exerts deflection-dependent variations in $n_m$, allowing us to express the nonlinear polarization $P_x^{NL}$ to the first order as[30]

$$P_x^{NL} = 2\varepsilon_0 n_m \left(\frac{\partial n_m}{\partial \delta_1}\delta_1 + \frac{\partial n_m}{\partial \delta_2}\delta_2\right) E_x,$$

where $n_m$ the modal refractive index, $\partial n_m/\partial \delta_j$ its change due to a deflection $\delta_j$ of the nanoweb $j$ and $E_x$ the transverse field component of the optical mode. Containing an infinite number of co-polarized equidistant-in-frequency components, it can be decomposed for slow and fast optical modes using the respective forms

$$E_x^s(x,y,z,t) = X_s(x)Y(y)\sqrt{\frac{P_0}{2\varepsilon_0 n_s^2 V_s}}\sum_n s_n(z)\exp\left[i(\beta_{s,n}z - \omega_n t)\right] + c.c.$$

$$E_x^f(x,y,z,t) = X_f(x)Y(y)\sqrt{\frac{P_0}{2\varepsilon_0 n_f^2 V_f}}\sum_n f_n(z)\exp\left[i(\beta_{f,n}z - \omega_n t)\right] + c.c. \qquad (5)$$

where $P_0$ is the launched optical power, $s_n$ and $f_n$ represent the slowly varying dimensionless field amplitudes of the $n^{th}$ comb lines (negative values correspond to Stokes components) for the slow and fast optical modes, $\omega_n = \omega_0 + n\Omega$, $\beta_{s,n} = \beta_{s,0} + nq$ and $\beta_{f,n} = \beta_{f,0} + nq$ are frequencies and axial propagation constants of the side-bands, $\omega_0$ is the optical pump frequency, $\beta_{s,0}$, $\beta_{f,0}$ are the corresponding wavevectors in slow and fast mode, $\Omega$ and $q$ frequency and propagation constant of the driving beat-note and $V_s$ and $V_f$ are the group velocities of the slow and fast optical modes. The transverse field distribution $X_l(x)Y(y)$ ($l = s$ or $f$) of the optical mode is normalized so that

$$\int_{-w_j/2}^{w_j/2} |X_l(x)|^2 \, dx = \int_{-\infty}^{\infty} |Y(y)|^2 \, dy = 1.$$

Note that this Ansatz allows us to express the power of each optical side-band in each mode in the simple forms $P_{s,n} = |s_n|^2 P_0$ and $P_{f,n} = |f_n|^2 P_0$.

Using Maxwell's stress tensor formalism, one can express the optical pressure as

$$p_j^{opt} = T_{yy}(y = y_j^{outer}) - T_{yy}(y = y_j^{inner}),$$

where $y_j^{outer,inner}$ stands for the upper and lower web boundaries. The only non-zero tensor component is written for TE-polarization as

$$T_{yy} = \frac{1}{2}\left[\varepsilon_0 E_x^2 - \mu_0\left(H_y^2 - H_z^2\right)\right]$$

with the vacuum permeability $\mu_0$.

The instantaneous deflection of the nanowebs due to flexural vibrations can be written as follows:

$$\delta_{jR}(x,z,t) = \delta_{jR0}(x)\sqrt{\frac{P_0}{2c\psi_{jR}\sigma\Omega_{jR}^2}} R_j(z)\exp[-i\Omega t]+\text{c.c.},\tag{6}$$

$$\delta_{jM}^{\pm}(x,z,t) = \delta_{jM0}(x)\sqrt{\frac{P_0}{2V_{jM}\psi_{jM}\sigma\Omega_{jM}^2}} M_j^{\pm}(z)\exp[i(\pm qz-\Omega t)]+\text{c.c.},\tag{7}$$

where the subscripts $R$ and $M$ stand for SRLS and SIMS phonons, $R_j(z)$ and $M_j^{\pm}(z)$ denote their dimensionless slowly-varying field envelopes ("+" for forward and "−" for backward phonons), $\Omega_{jR}$ and $\Omega_{jM}$ the phonon eigenfrequencies, $V_{jM}$ the group velocity of the SIMS phonons (the group velocity of the SRLS phonons is negligible), $\delta_{jR0}(x)$ and $\delta_{jM0}(x)$ are the transverse profiles of the SRLS and SIMS phonons (profiles for forward and backward phonons are supposed to be identical), normalized as $\int_{-w_j/2}^{w_j/2} |\delta_{jR0}(x)|^2 \, dx = \int_{-w_j/2}^{w_j/2} |\delta_{jM0}(x)|^2 \, dx = 1$. The dimensionless normalization coefficients $\psi_{jR}$ and $\psi_{jM}$ are defined as

$$\psi_{jR} = \frac{D_j}{\sigma_j \Omega_{jR}^2} \int_{-w_j/2}^{w_j/2} \left[\frac{\partial^2 \delta_{jR0}}{\partial x^2}\right]^2 dx,$$

$$\psi_{jM} = \frac{D_j}{\sigma_j \Omega_{jM}^2} \left( \int_{-w_j/2}^{w_j/2} \left( \left[\frac{\partial^2 \delta_{jM0}}{\partial x^2}\right]^2 - 2q_{jM}^2 \left(\delta_{jM0} \frac{\partial^2 \delta_{jM0}}{\partial x^2} + (1-\nu)\left[\frac{\partial \delta_{jM0}}{\partial x}\right]^2\right)\right) dx + q_{jM}^4 \right),$$

so that the acoustic energy per unit length for SRLS vibrations $e_R$ and the power carried by SIMS phonons $P_M$ can be expressed as $e_R = (P_0/c)\sum_j |R_j|^2$ and $P_M = P_0 \sum_j |M_j^\pm|^2$ [41].

Firstly, we obtain an eigenvalue equation for flexural modes by inserting Eq.(7) into Eq.(3), while dropping the dependence of the envelopes on axial position and setting optical pressure to zero:

$$\left[\frac{\partial^4}{\partial x^4} - 2q^2 \frac{\partial^2}{\partial x^2}\right] \delta_{jM0}(x) = \left[q_0^4 - q^4\right] \delta_{jM0}(x),$$

where $q_0 = (\sigma \Omega^2 / D)^{1/4}$ is the transverse wavenumber. The symmetric general solution of this equation can be written as[40]

$$\delta_{jM0}(x) = C_1 \cosh\left(\sqrt{q^2 - q_0^2}\, x\right) + C_2 \cosh\left(\sqrt{q^2 + q_0^2}\, x\right), \tag{8}$$

where $C_{1,2}$ are constants. Fixed-edge boundary conditions $\delta_{jM0}(x = \pm w_j/2) = \partial_x \delta_{jM0}(x = \pm w_j/2) = 0$ yield the following dispersion relation of the flexural vibrations:

$$\sqrt{q^2 - q_0^2}\, \tanh\left(w_j/2 \sqrt{q^2 - q_0^2}\right) = \sqrt{q^2 + q_0^2}\, \tanh\left(w_j/2 \sqrt{q^2 + q_0^2}\right). \tag{9}$$

This equation shows that the spacing between the flexural dispersion curves in opposite nanowebs is tunable by adjusting the nanoweb widths. The characteristic dispersion diagrams for flexural modes are presented in Fig. 2(a). The phase-matching for SRLS transitions is met at cut-off frequencies. Therefore, both the transverse shape and the eigenfrequency of the SRLS vibrations can be found from Eqs.(8) and (9) for $q = 0$.

While a slight change in thickness along the nanowebs will not affect the flexural vibrations, this has a huge impact on the optical modes: bound ones with low loss that do not penetrate into the fiber cladding can only be obtained at positions along the nanowebs where their thickness is locally increased. To simulate their effective index and transverse shape, we perform finite-element modelling using the commercial software COMSOL Multiphysics and assume both nanowebs to feature the same convex Gaussian thickness profile, defined by its waist $\Delta x$ and the change in thickness $\Delta h$. With this, the coordinates of the glass-air boundaries inside and outside the interweb gap become $y_j^{outer,inner} = \pm(h_g + h)/2 \pm \{h/2 + \Delta h \exp(-[x - w_j/2]^2 / \Delta x^2)\}$, respectively. For our simulations, we extract typical values $\Delta h = 100$ nm and $\Delta x = 1.5$ μm from a scanning electron micrograph of the experimental dual-nanoweb structure[27]. Although basically SIMS transitions are possible between any pair of co-propagating optical modes, their gain is maximum for transitions between the fundamental and the respective higher-order mode closest in wavevector. This is because an increasing SIMS wavevector $q$ causes narrowing of the transverse shape of the vibrational mode (see Eq.(8)), at the same time decreasing the overlap between optical and acoustic modes. Simulations have shown that the closest wavevector with respect to the fundamental optical mode is met for the mode with a double-lobed transverse field distribution along the $x$-coordinate (see Fig. 1(c)).

Next, we evaluate the dephasing for transitions between adjacent sidebands. At a wavelength of 1.55 μm and for a web width of ~20 μm the typical modal indices are $n_s = 1.309$ and $n_f = 1.275$ for slow and fast modes, respectively. The cut-off frequencies of flexural vibrations are $\Omega \sim 2\pi \times 5$ MHz, resulting in the beat-note SRLS and SIMS wavevectors $q_R = \Omega n_s / c \approx 0.14$ m$^{-1}$ and $q_M = (n_s \omega_0 - n_f \omega_{-1})/c \approx 140$ mm$^{-1}$. These translate to acoustic wavelengths of ~45 m (SRLS) and ~45 μm (SIMS). A typical group velocity dispersion $\beta_2 = 500$ ps$^2$/km yields enormous characteristic dephasing lengths: it is $\pi/(\beta_2 \Omega^2) \sim 10^6$ km for SRLS transitions, whereas for SIMS transitions it is $2\pi c/((n_s - n_f)\Omega) \sim 22$ km. Hence, we can neglect the dephasing between adjacent comb lines for both SRLS and SIMS phonons as typical fibre lengths used in experiments are orders of magnitude below these limits.

Taking into account all these considerations, we derive the steady-state equations (1a)–(1e) governing the slowly-varying field amplitudes of each optical side-band and the flexural envelopes. To this end, Eqs.(5), (6) and (7) are substituted into the wave equations (Eqs.(3) and (4)) while using Eqs.(8) and (9). The rates of opto-acoustic coupling for the various transitions are given by

$$\kappa^j_{Rmm} = \frac{P_0 Q_{jRmm} H_{jRmm}}{4c\Omega \sigma n_m} \sqrt{\frac{2c\psi_{jR}\sigma\Omega^2_{jR}}{P_0}}, \quad \kappa^j_{Msf} = \frac{P_0 Q_{jMsf} H_{jMsf}}{4c\Omega \sigma \sqrt{n_s n_f}} \sqrt{\frac{2V_{jM}\psi_{jM}\sigma\Omega^2_{jM}}{P_0}},$$

$$\kappa^j_{mRm} = \frac{\omega_0}{c}\frac{\partial n_m}{\partial \delta_j} Q_{jRmm} \sqrt{\frac{P_0}{2c\psi_{jR}\sigma\Omega^2_{jR}}}, \quad \kappa^j_{sMf} = \sqrt{\frac{n_f}{n_s}}\frac{\omega_0}{c}\frac{\partial n_f}{\partial \delta_j} Q_{jMsf} \sqrt{\frac{P_0}{2V_{jM}\psi_{jM}\sigma\Omega^2_{jM}}},$$

$$\kappa^j_{fMs} = \sqrt{\frac{n_s}{n_f}}\frac{\omega_0}{c}\frac{\partial n_s}{\partial \delta_j} Q_{jMsf} \sqrt{\frac{P_0}{2V_{jM}\psi_{jM}\sigma\Omega^2_{jM}}},$$

where $m = s$ or $f$ stands for slow and fast optical modes. The Brillouin linewidths of SRLS and SIMS phonons are $\Gamma_{jR} = \tau\Omega^2_{jR}$ and $\Gamma_{jM} = \tau\Omega^2_{jM}$, respectively. The optomechanical overlap integrals of the transverse mode shapes involved in SRLS and SIMS are defined as

$$Q_{jRmm} = \int_{-w_j/2}^{w_j/2} \delta_{jR0}(x)\,|X_m(x)|^2\,dx, \quad Q_{jMsf} = \int_{-w_j/2}^{w_j/2} \delta_{jM0}(x)\,|X_s(x)||X_f(x)|\,dx.$$

The quantities

$$H_{jRmm} = \left[(1-n_m^2)Y^2 - \left(\frac{c}{\omega_0}\right)^2\left|\frac{\partial Y}{\partial y}\right|^2\right]_{y=y_j^{lower}}^{y=y_j^{upper}}, \quad H_{jMsf} = \left[(1-n_s n_f)Y^2 - \left(\frac{c}{\omega_0}\right)^2\left|\frac{\partial Y}{\partial y}\right|^2\right]_{y=y_j^{inner}}^{y=y_j^{outer}}$$

describe coupling of the optical fields to the flexural vibrations for the SRLS and SIMS transitions. In Eqs.(1d) and (1e), the group velocity of SIMS phonons appears as

$$V_{jM} = 2q\frac{D}{\sigma\Omega_{jM}}\left|\int_{-w_j/2}^{w_j/2}\delta_{jM0}\frac{\partial^2\delta_{jM0}}{\partial x^2}dx\right|.$$

**Estimations of typical parameters for a dual-web fibre.**
To evaluate all quantities defined before, we numerically calculate the transverse profiles $X_m(x)Y(y)$ and dispersion properties of the fast and slow optical modes for parameters $h_g = h = 500$ nm, $w_1 = 23$ μm, $w_2 = 28.06$ μm, $\Delta h = 100$ nm and $\Delta x = 1.5$ μm at an optical pump wavelength $\lambda = 1.55$ μm. The results yielded $H_{jRss} \approx -0.16$ μm$^{-1}$, $H_{jRff} \approx -0.154$ μm$^{-1}$ and $H_{jMsf} \approx -0.156$ μm$^{-1}$. Variations in $n_s$ and $n_f$ induced by deflection of the nanowebs result in the following nonlinear optomechanical coefficients $\partial n_s/\partial\delta_j = -0.058$ μm$^{-1}$ and $\partial n_f/\partial\delta_j = -0.059$ μm$^{-1}$. Using elastic parameters of silica $\rho = 2.2\times10^3$ kg/m$^3$, $\Upsilon = 75.2\times10^9$ N/m$^2$ and $\nu = 0.17$, we estimate $\Omega_0 = 2\pi\times5.6$ MHz, $V_{1M} = V_{2M} = 12$ m/s, $Q_{jRss} \approx 320$ m$^{-1/2}$, $Q_{jRff} \approx 300$ m$^{-1/2}$ and $Q_{jMsf} \approx 245$ m$^{-1/2}$. The Brillouin linewidths $\Gamma_{jR}/2\pi = 400$ Hz and $\Gamma_{jM}/2\pi = 800$ Hz were extracted from experimental measurements[18] and lead to an effective optical noise power $P_{noise} \approx 0.2$ nW.

**REFERENCES**

(1) Butsch, A.; Conti, C.; Biancalana, F.; Russell, P. S. J. Optomechanical Self-Channeling of Light in a Suspended Planar Dual-Nanoweb Waveguide. *Phys. Rev. Lett.* **2012**, *108*, 093903.
(2) Butsch, A.; Kang, M. S.; Euser, T. G.; Koehler, J. R.; Rammler, S.; Keding, R.; Russell, P. S. J. Optomechanical Nonlinearity in Dual-Nanoweb Structure Suspended Inside Capillary Fiber. *Phys. Rev. Lett.* **2012**, *109*, 183904.
(3) Shin, H.; Qiu, W.; Jarecki, R.; Cox, J. A.; Iii, R. H. O.; Starbuck, A.; Wang, Z.; Rakich, P. T. Tailorable Stimulated Brillouin Scattering in Nanoscale Silicon Waveguides. *Nat. Commun.* **2013**, *4*, 1944.
(4) Van Laer, R.; Kuyken, B.; Van Thourhout, D.; Baets, R. Interaction between Light and Highly Confined Hypersound in a Silicon Photonic Nanowire. *Nat. Photonics* **2015**, *9*, 199–203.
(5) Rakich, P. T.; Reinke, C.; Camacho, R.; Davids, P.; Wang, Z. Giant Enhancement of Stimulated Brillouin Scattering in the Subwavelength Limit. *Phys. Rev. X* **2012**, *2*, 011008.

(6) Vidal, B.; Piqueras, M. A.; Martí, J. Tunable and Reconfigurable Photonic Microwave Filter Based on Stimulated Brillouin Scattering. *Opt. Lett.* **2007**, *32*, 23–25.
(7) Zhang, W.; Minasian, R. A. Ultrawide Tunable Microwave Photonic Notch Filter Based on Stimulated Brillouin Scattering. *IEEE Photonics Technol. Lett.* **2012**, *24*, 1182–1184.
(8) Li, J.; Lee, H.; Vahala, K. J. Microwave Synthesizer Using an On-Chip Brillouin Oscillator. *Nat. Commun.* **2013**, *4*, 3097.
(9) Marpaung, D.; Morrison, B.; Pagani, M.; Pant, R.; Choi, D.-Y.; Luther-Davies, B.; Madden, S. J.; Eggleton, B. J. Low-Power, Chip-Based Stimulated Brillouin Scattering Microwave Photonic Filter with Ultrahigh Selectivity. *Optica* **2015**, *2*, 76–83.
(10) Shin, H.; Cox, J. A.; Jarecki, R.; Starbuck, A.; Wang, Z.; Rakich, P. T. Control of Coherent Information via On-Chip Photonic–phononic Emitter–receivers. *Nat. Commun.* **2015**, *6*, 6427.
(11) Braje, D.; Hollberg, L.; Diddams, S. Brillouin-Enhanced Hyperparametric Generation of an Optical Frequency Comb in a Monolithic Highly Nonlinear Fiber Cavity Pumped by a Cw Laser. *Phys. Rev. Lett.* **2009**, *102*, 193902.
(12) Dong, M.; Winful, H. G. Unified Approach to Cascaded Stimulated Brillouin Scattering and Frequency-Comb Generation. *Phys. Rev. A* **2016**, *93*, 043851.
(13) Büttner, T. F. S.; Kabakova, I. V.; Hudson, D. D.; Pant, R.; Poulton, C. G.; Judge, A. C.; Eggleton, B. J. Phase-Locking and Pulse Generation in Multi-Frequency Brillouin Oscillator via Four Wave Mixing. *Sci. Rep.* **2014**, *4*, 05032.
(14) Li, J.; Lee, H.; Vahala, K. J. Low-Noise Brillouin Laser on a Chip at 1064 Nm. *Opt. Lett.* **2014**, *39*, 287–290.
(15) Eggleton, B. J.; Poulton, C. G.; Pant, R. Inducing and Harnessing Stimulated Brillouin Scattering in Photonic Integrated Circuits. *Adv. Opt. Photonics* **2013**, *5*, 536–587.
(16) Kang, M. S.; Brenn, A.; St.J. Russell, P. All-Optical Control of Gigahertz Acoustic Resonances by Forward Stimulated Interpolarization Scattering in a Photonic Crystal Fiber. *Phys. Rev. Lett.* **2010**, *105*, 153901.
(17) Kittlaus, E. A.; Otterstrom, N. T.; Rakich, P. T. On-Chip Inter-Modal Brillouin Scattering. *Nat. Commun.* **2017**, *8*, 15819.
(18) Koehler, J. R.; Noskov, R. E.; Sukhorukov, A. A.; Butsch, A.; Novoa, D.; Russell, P. S. J. Resolving the Mystery of Milliwatt-Threshold Opto-Mechanical Self-Oscillation in Dual-Nanoweb Fiber. *APL Photonics* **2016**, *1*, 056101.
(19) Kang, M. S.; Butsch, A.; Russell, P. S. J. Reconfigurable Light-Driven Opto-Acoustic Isolators in Photonic Crystal Fibre. *Nat. Photonics* **2011**, *5*, 549–553.
(20) Kang, M. S.; Nazarkin, A.; Brenn, A.; Russell, P. S. J. Tightly Trapped Acoustic Phonons in Photonic Crystal Fibres as Highly Nonlinear Artificial Raman Oscillators. *Nat. Phys.* **2009**, *5*, 276–280.
(21) Shen, Y. R.; Bloembergen, N. Theory of Stimulated Brillouin and Raman Scattering. *Phys. Rev.* **1965**, *137*, A1787–A1805.
(22) Duncan, M. D.; Mahon, R.; Reintjes, J.; Tankersley, L. L. Parametric Raman Gain Suppression in D2 and H2. *Opt. Lett.* **1986**, *11*, 803–805.
(23) Bauerschmidt, S. T.; Novoa, D.; Russell, P. S. J. Dramatic Raman Gain Suppression in the Vicinity of the Zero Dispersion Point in a Gas-Filled Hollow-Core Photonic Crystal Fiber. *Phys. Rev. Lett.* **2015**, *115*, 243901.

(24) Hosseini, P.; Mridha, M. K.; Novoa, D.; Abdolvand, A.; Russell, P. S. J. Universality of Coherent Raman Gain Suppression in Gas-Filled Broadband-Guiding Photonic Crystal Fibers. *Phys. Rev. Appl.* **2017**, *7*, 034021.

(25) Dainese, P.; Russell, P. S. J.; Wiederhecker, G. S.; Joly, N.; Fragnito, H. L.; Laude, V.; Khelif, A. Raman-like Light Scattering from Acoustic Phonons in Photonic Crystal Fiber. *Opt. Express* **2006**, *14*, 4141–4150.

(26) Kittlaus, E. A.; Shin, H.; Rakich, P. T. Large Brillouin Amplification in Silicon. *Nat. Photonics* **2016**, *10*, 463.

(27) Butsch, A.; Koehler, J. R.; Noskov, R. E.; Russell, P. S. J. CW-Pumped Single-Pass Frequency Comb Generation by Resonant Optomechanical Nonlinearity in Dual-Nanoweb Fiber. *Optica* **2014**, *1*, 158–164.

(28) Van Laer, R.; Baets, R.; Van Thourhout, D. Unifying Brillouin Scattering and Cavity Optomechanics. *Phys. Rev. A* **2016**, *93*, 053828.

(29) Wolff, C.; Steel, M. J.; Eggleton, B. J.; Poulton, C. G. Stimulated Brillouin Scattering in Integrated Photonic Waveguides: Forces, Scattering Mechanisms, and Coupled-Mode Analysis. *Phys. Rev. A* **2015**, *92*, 013836.

(30) Boyd, R. W. *Nonlinear Optics, Third Edition*, 3 edition.; Academic Press: Amsterdam ; Boston, 2008.

(31) Kobyakov, A.; Sauer, M.; Chowdhury, D. Stimulated Brillouin Scattering in Optical Fibers. *Adv. Opt. Photonics* **2010**, *2*, 1.

(32) Pant, R.; Poulton, C. G.; Choi, D.-Y.; Mcfarlane, H.; Hile, S.; Li, E.; Thevenaz, L.; Luther-Davies, B.; Madden, S. J.; Eggleton, B. J. On-Chip Stimulated Brillouin Scattering. *Opt. Express* **2011**, *19*, 8285.

(33) Koehler, J. R.; Noskov, R. E.; Sukhorukov, A. A.; Novoa, D.; Russell, P. S. J. Coherent Control of Flexural Vibrations in Dual-Nanoweb Fiber Using Phase-Modulated Two-Frequency Light. *ArXiv170607311 Phys.* **2017**.

(34) Dai, D.; Wang, J.; Shi, Y. Silicon Mode (de)Multiplexer Enabling High Capacity Photonic Networks-on-Chip with a Single-Wavelength-Carrier Light. *Opt. Lett.* **2013**, *38*, 1422–1424.

(35) Luo, L.-W.; Ophir, N.; Chen, C. P.; Gabrielli, L. H.; Poitras, C. B.; Bergmen, K.; Lipson, M. WDM-Compatible Mode-Division Multiplexing on a Silicon Chip. *Nat. Commun.* **2014**, *5*, 4069.

(36) Yao, J. Microwave Photonics. *J. Light. Technol.* **2009**, *27*, 314–335.

(37) Marpaung, D.; Roeloffzen, C.; Heideman, R.; Leinse, A.; Sales, S.; Capmany, J. Integrated Microwave Photonics. *Laser Photonics Rev.* **2013**, *7*, 506–538.

(38) Diddams, S. A. The Evolving Optical Frequency Comb. *JOSA B* **2010**, *27*, B51–B62.

(39) Landau, L. D.; Pitaevskii, L. P.; Kosevich, A. M.; Lifshitz, E. M. *Theory of Elasticity, Third Edition: Volume 7*, 3 edition.; Butterworth-Heinemann, 1986.

(40) Norris, A. N. Flexural Waves on Narrow Plates. *J. Acoust. Soc. Am.* **2003**, *113*, 2647–2658.

(41) Szilard, R. *Theories and Applications of Plate Analysis: Classical, Numerical and Engineering Methods*; John Wiley & Sons: Hoboken, New Jersey, 2004.